\documentclass[aps,onecolumn,showpacs,showkeys,nofootinbib]{revtex4}
\usepackage{epsfig}
\usepackage{amsmath}
\usepackage{amsfonts}
\usepackage{amssymb}
\usepackage{graphicx}
\usepackage{colordvi}
\usepackage{color}
\usepackage{graphics}
\usepackage{bm}

\begin{document}

\title{The gravitino problem in Extended Gravity cosmologies}

\author{Salvatore Capozziello$^{a,b,c,d}$\footnote{e-mail address: capozziello@na.infn.it} and
Gaetano Lambiase$^{e,f}$\footnote{e-mail address: lambiase@sa.infn.it} }

\affiliation{$^a$Dipartimento di Fisica  ``E. Pancini", Universit\`a di Napoli  ``Federico II", Via Cinthia, I-80126, Napoli, Italy.}
\affiliation{$^b$Istituto Nazionale di Fisica Nucleare (INFN), sez. di Napoli, Via Cinthia 9, I-80126 Napoli, Italy.}
\affiliation{$^c$Scuola Superiore Meridionale, Largo San Marcellino 10, 80138 Napoli, Italy.}
\affiliation{$^d$Laboratory for Theoretical Cosmology, Tomsk State University of Control Systems and Radioelectronics (TUSUR), 634050 Tomsk,  Russia. }
\affiliation{$^e$Dipartimento di Fisica E.R. Cainaiello, University of Salerno, Via Giovanni Paolo II, I 84084-Fisciano (SA), Italy.}
\affiliation{$^f$INFN, Gruppo Collegato di Salerno, Sezione di Napoli, Via Giovanni Paolo II, I 84084-Fisciano (SA), Italy.}



\begin{abstract}
The gravitino problem is investigated in the framework of Extended Gravity  cosmologies. 
In particular, we consider $f(R)$ gravity, the most natural extension of the Hilbert-Einstein action, and $f(\cal T)$ gravity, the extension of teleparallel equivalent gravity. Since
in these theories the expansion laws of the Universe are modified, as compared to the standard $\Lambda$CDM cosmology, it follows that also the thermal history of particles gets modified. We show that $f(R)$ models allow to avoid the
late abundance of gravitinos. In particular, we found that for an appropriate choice of the parameters characterizing the $f(R)$ model, the gravitino abundance turns out to be independent of the reheating temperature. A similar behavior is achieved also in the context of $f(\cal T)$ gravity. In this perspective, we can conclude that geometric corrections to standard General Relativity (and to Teleparallel Equivalent of General Relativity) can improve shortcomings both in cosmology and in unified theories beyond the Standard Model of particles.
\end{abstract}



\maketitle

\section{Introduction}

One of the most attractive extension of the Standard Model of Particles (SM) is represented by  Supersymmetry (SUSY) \cite{wess} since it provides the stability of electroweak scale against radiative corrections, and since the three gauge couplings of the SM meet at GUT scales $10^{16}$GeV. Apart these motivations, there are, up to now,  no direct evidence of the existence of superpartner particles predicted by the theory. A natural extension of SUSY is the Supergravity (SUGRA), which is essentially the local counterpart of SUSY \cite{ferrara}. The SUGRA models predict the existence of  the {\it gravitino} (the superpartner of the graviton). The latter, in some SUSY models \cite{ellisN}, is characterized by the fact that its mass is not directly related to the SUSY breaking scales of SM ordinary particles  and their superpartners. It is worth noticing  that its interaction is weak, so that the chance to find it in collider experiments is extremely unlikely. On the other hand, the gravitino can have interesting consequences at cosmological scales, in particular during the early phases of the Universe evolution.

An open issue in cosmology is the so called {\it gravitino problem} \cite{sat,gravitinoPr}. Since SUSY particles couple to (ordinary) matter only through the gravitational interaction, it turns out that their couplings are Planck suppressed, implying a quite long lifetime
\begin{equation}
 \tau\sim \frac{M_{Pl}^2}{m_{3/2}}\simeq 10^5\left(\frac{1\mbox{TeV}}{m_{3/2}}\right)^3 \mbox{sec}\,,
  \end{equation}
where $m_{3/2}$ is the gravitino mass. The scale characterizing $m_{3/2}$ is of the order of electroweak energies ${\cal O}10^2$GeV. One of the consequences of particles with so long lifetime is that, if they decay after the Big Bang nucleosynthesis (BBN), then their decaying  products would destroy light elements, and hence in contradiction with the successful predictions of BBN. This problem is avoided by setting an upper bound on the reheating temperature. In the framework of General Relativity (GR), the gravitino abundance $Y_{3/2}$ is proportional to the reheating temperature $T_R$ \cite{moroi}
 \begin{equation}\label{Yabundance}
 Y_{3/2} \simeq 10^{-11}\frac{T_R}{10^{10}\mbox{GeV}}\,,
 \end{equation}
so that requiring that it remains small $Y_{3/2}\lesssim 10^{-14}$ for a successful prediction of BBN, one gets the constraint \cite{olive}
 \begin{equation}\label{upperTR}
T_R \lesssim (10^6 -10^7)\text{GeV} \quad \mbox{for} \quad m_{3/2} \sim {\cal O}(10^2\text{GeV})\,.
 \end{equation}
The bound (\ref{upperTR}) opens some issues related to the inflationary scenarios since the latter provides a reheating temperature larger than the upper bound (\ref{upperTR})  (see  \cite{linde,mazumdar}).

In this paper,  we consider the gravitino problem in the framework Extended Gravity cosmologies.  As possible models of Extended  Gravity, we shall consider $f(R)$ theory, which represents the simplest extension of theories involving  the Ricci curvature scalar $R$, and the $f({\cal T})$ theory, where ${\cal T}$ is the torsion field adopted in the Teleparallel Equivalent of General Relativity (TEGR). These theories have been recently  invoked in order to  explain the  discovery of  accelerated expansion of the Universe \cite{accUn}, without introducing unknown forms of Dark Energy, and, in the astrophsyical contexts, to fix the  Dark Matter problem. The appearance of these  dark components in the Universe dynamics is a clear signal of the breakdown of GR at infrared   scales.

The aim of this paper is to show that, considering cosmological models related to these theories can contribute to alleviate the above gravitino problem.

The paper is organized as follows. In Section II, we discuss the gravitino problem in the framework of $f(R)$ cosmology. The  $f({\cal T})$ cosmology  and the gravitino problem is considered  in Sec. III.  Discussion and conclusions are  reported  in Sec. IV.

\section{The gravitino problem in $f(R)$ cosmology}

The gravitational action for $f(R)$ gravity is \cite{amendolabook,odintsovreport,curvquin}
\begin{equation}\label{Lagr}
  S=\frac{1}{2\kappa^2}\int d^4x \sqrt{-g}\, f(R)+S_m[g_{\mu\nu},
  \psi_m]\,,
\end{equation}
where $S_m$ is the matter action and $\kappa^2=8\pi G=8\pi/M_{Pl}^2$ is the gravitational coupling ($M_{Pl}\simeq 10^{19}$GeV is the Planck mass).
The action (\ref{Lagr}) must be considered as an effective theory that allows to describe, at a phenomenological level, the gravitational interactions.
The field equations for $f(R)$ gravity are
\begin{equation}\label{fieldeqs}
  f' R_{\mu\nu}-\frac{f}{2}\, g_{\mu\nu}-\nabla_\mu \nabla_\nu f' +g_{\mu\nu}\Box f'= \kappa^2 T^m_{\mu\nu}\,,
\end{equation}
while the trace equation is given by
\begin{equation}\label{tracef}
  3\Box f'+f' R-2f=\kappa^2 T^m\,, \qquad T^m = \rho-3p\,.
\end{equation}
Here $f^\prime\equiv \displaystyle{\frac{\partial f}{\partial R}}$. In cosmology, it is  $\Box f'= {\ddot f}^\prime+3H{\dot f}^{\prime}$, $H={\dot a}/a$, and the dot stands for $d/dt$. $T^m_{\mu\nu}$ is the matter energy-momentum tensor  and $T^m$ its trace.  The tensor $T^m_{\mu\nu}$ is divergenceless, i.e.  $\nabla^\mu T^m_{\mu\nu}=0$ and this property holds also for the l.h.s. of \eqref{fieldeqs}.
There are several hypotheses on the explicit
form of  function $f(R)$, however, at the moment, there is no final statement on a comprehensive  form working at any scale. For the sake of simplicity,  we adopt a power law model (for other models and applications, see \cite{amendolabook,odintsovreport,curvquin,fornengo})
 \begin{equation}\label{f(R)=R+aR**ng}
    f(R) = \alpha R^n\,,
 \end{equation}
 where $n$ is a real number. Clearly, standard GR is recovered for $n=1$.

In a (spatially flat) Friedman-Lema\^itre-Robertson-Walker (FLRW) metric $ ds^2=dt^2-a^2(t)[dx^2+dy^2+dz^2]$, with the ansatz $a(t)=a_0 t^\beta$, the $0-0$ field equation and the trace equation read
 \begin{eqnarray}\label{Hmodified0}
    \frac{\alpha}{2}\left[\frac{n(\beta+2n-3)}{2\beta-1}-1\right]R^n &=& \kappa^2 \rho\,, \\
   \alpha \left[n-2-\frac{n(n-1)(2n-1)}{\beta(1-2\beta)}+\frac{3n(n-1)}{1-2\beta}\right] R^n &=& \kappa^2 (1-3w)\rho\,, \label{Hmodifiedtrace0}
 \end{eqnarray}
where $\rho$ is the energy density which, in the radiation dominated era,  reads 
\begin{equation}
\rho^m=\frac{\pi^2 g_*}{30}T^4\,,
 \end{equation}
 where $g_*\simeq 10^2$ counts the number of relativistic degrees of freedom.

 In a  radiation dominated Universe, where the trace $T^m=0$, Eqs. (\ref{Hmodified0}) and (\ref{Hmodifiedtrace0}) yield
 \begin{equation}\label{Hmodified}
    \alpha \Omega_n R^n = \kappa^2 \rho\,,  
 \end{equation}
where we have defined de quantities
 \begin{equation}\label{Omega}
   \Omega_{n} \equiv \frac{5n^2-8n+2}{4(n-1)}\, \quad \beta = \frac{n}{2}\,.
 \end{equation}
From Eq. (\ref{Hmodified}),  one obtains the relation between the cosmic time $t$ and the temperature $T$
 \begin{equation}\label{t-T}
   t = \Pi_n \left(\frac{T}{M_{Pl}}\right)^{-\frac{2}{n}} M_{Pl}^{-1}\,,
 \end{equation}
here, it is 
 \begin{equation}\label{Pi}
   \Pi_n \equiv [3n|n-1|]^{1/2}\left(\frac{15{\tilde \alpha}\Omega_{n}}{4\pi^3 g_*}\right)^{\frac{1}{2n}}\,, \qquad
   {\tilde \alpha}=\frac{\alpha}{M_{Pl}^{2(1-n)}}\,.
 \end{equation}
Let us introduce the transition time (temperature) $t_*$ ($T_*$) which characterizes the transition
from the $f(R)$ cosmology to the standard cosmology, described by GR. This means to equate the equation of the evolution at the instant $t=t_*$, i.e. $\alpha \Omega_{\beta,n}R^n(t_*)=H_{GR}^2(t_*)$. One gets
\begin{equation}\label{t-transition}
    t_* = \left[4{\tilde \alpha}\Omega_{n}[3n|n-1|]^n\right]^{\frac{1}{2(n-1)}}M_{Pl}^{-1}\,,
\end{equation}
The expression of the transition temperature $T_*$ is given by
  \begin{equation}\label{Ttransition}
    T_* \equiv M_{Pl} [3n|n-1|]^{-\frac{n}{4(n-1)}}\left[\frac{15}{16\pi^3 g_*}\right]^{\frac{1}{4}}
    \left[4{\tilde \alpha}\Omega_{n}\right]^{-\frac{1}{4(n-1)}}\,,
 \end{equation}
so that  relation (\ref{t-transition}) can be cast in the form
\begin{equation}\label{t-transition2}
  t = t_* \left(\frac{T}{T_*}\right)^{-\frac{2}{n}} \,.
\end{equation}
Moreover, notice that
 \begin{equation}\label{t*T*}
   \frac{t_* T_*^2}{M_{Pl}}=\sqrt{\frac{15}{16\pi^3 g_*}}\,.
 \end{equation}
In what follows, we shall assume that, after inflation, from GUT scales to reheating scales,  the Universe evolves according to (\ref{f(R)=R+aR**ng}), then its evolution  enters in the regime governed by the cosmological standard model. Hence, the $f(R)$ cosmology provides a scenario in which the thermal history of particles gets modified, and as a consequence, the Boltzmann equation too (see below). This reflects on the gravitino abundance, which turns out to depend on parameters characterizing the $f(R)$ model and the transition temperature, i.e. the temperature at which the Universe passes from the $f(R)$ evolution to the standard GR one. In this phase, the parameters $\{\alpha, n\}$ can play an important role.

Let us  notice that the expansion rate of the Universe in $f(R)$ cosmology can be written as
\begin{eqnarray}
    H(T) &=& A(T)\, H^{(GR)}(T)\,, \label{ATf} \\
    A(T) &\equiv& \eta \left(\frac{T}{T_*}\right)^p  \label{ATGen} \\
   \mbox{with} \qquad \eta &\equiv & 2\sqrt{3}\beta \,, \nonumber \\
     p &\equiv & \frac{2}{n}-2  \nonumber
\end{eqnarray}
where  $A(T)$ is an enhancement factor. Expressions
similar to (\ref{ATf}) are obtained in different frameworks: $p=2$ in Randall-Sundrum type II brane cosmology \cite{randal}, $p=1$ in kination models \cite{kination}, $p=-1$ in scalar-tensor cosmology \cite{STcosmology},
$-1\lesssim p \lesssim 0$ in various $f(R)$ cosmologies \cite{PRDGall}.


As pointed out in the Introduction, gravitino is generated by means of thermal scattering in the primordial plasma.
This occurs during the reheating era after inflation. To describe the gravitino production,  one makes use of
the Boltzmann equation for the number density of species in thermal bath. 
Following Ref. \cite{sat}, the relevant equation for gravitino production is 
 \begin{equation}\label{boltzmann1}
   \frac{d n_{3/2}}{dt}+3Hn_{3/2}=\langle \sigma v \rangle n_{\text{rad}}^2\,.
 \end{equation}
Here $n_{3/2, \text{rad}}$ refers to gravitino and relativistic species, while $\langle \ldots  \rangle$
stands for the thermal average of the gravitino cross section $\sigma v$  times the relative velocity
of scattering radiation ($v \sim 1$). In (\ref{boltzmann1}),  we have neglected the term
$\displaystyle{\frac{m_{3/2}}{\langle E_{3/2}\rangle} \frac{n_{3/2}}{\tau_{3/2}}}$, where $\displaystyle{\frac{m_{3/2}}{\langle E_{3/2}\rangle}}$ is the average Lorentz factor. Introducing the gravitino and relativistic particles abundances $Y_{3/2}=n_{3/2}/s$ and $Y_{\text{rad}}=n_{\text{rad}}/s$, respectively, where $s=\frac{2\pi^2}{45}\, g_* T^3$
and $g_*\sim 300$, the Boltzmann Eq. (\ref{boltzmann1}) assumes the form
 \begin{equation}\label{boltzmann2}
   \frac{dY_{3/2}}{dT}=\frac{ s \langle \sigma v \rangle}{{\dot T}} Y_{\text{rad}}^2\,.
 \end{equation}
Here we  used
 \begin{equation}
 \frac{\dot T}{T}=-\frac{n}{2t}=-\frac{n}{2t_*} \left(\frac{T}{T_*}\right)^{\frac{2}{n}}\,.
 \end{equation}
Integrating from $T_R$ ($\gg T_*$, see below) up to a low temperature $T_{l}$ ($\lesssim T_*$) in the era described by GR, the solution to (\ref{boltzmann2}) is

 \begin{equation}\label{Yfcosmology}
   Y_{3/2}\simeq -{\cal B}\frac{M_{Pl}}{T_*}\, \left[\left(\frac{T_R}{T_*}\right)^\Delta -\left(\frac{T_l}{T_*}\right)^\Delta\right]\,,
 \end{equation}
where
 \begin{eqnarray}
   \Delta & \equiv & 3-\frac{2}{n}\,, \label{Delta}\\
   {\cal B} &\equiv & \left[M_{Pl}\frac{\langle \sigma v\rangle s}{H_{GR} T} Y_{\text{rad}}^2\right]_R
    \frac{1}{\sqrt{3}(3n-2)}\,. \label{calB}
 \end{eqnarray}
Notice that (\ref{Yfcosmology}) is independent of the parameter $\alpha$.

The gravitino abundance derived in  Eq. (\ref{Yfcosmology}) allows to alleviate the gravitino problem. In fact,
for $\Delta \approx 0$, i.e. $n\approx 2/3$, it follows that the gravitino abundance turns out to be $Y_{3/2} << 1$. The most stringent constrain on unstable massive relic particles with lifetime $\gtrsim 10^2$sec, obtained from ${}^6\text{Li}$ abundance, is $Y_{3/2} \lesssim 10^{-14}\displaystyle{\frac{10^2\text{GeV}}{m_{3/2}}}$ \cite{olive}. Using this value it follows that the overproduction of gravitino is avoided if the transition temperature  is $T_* \lesssim (10^{6}-10^7)$GeV, while the reheating temperature can be larger in order to be compatible with inflationary prediction values.





\section{The gravitino problem in $f({\cal T})$ cosmology}

An interesting approach to gravity  is the TEGR,  firstly proposed by Einstein himself  \cite{Ferraro2,Pereira.book}. This theory of
gravity is based on the Weitzenb\"{o}ck connection (instead of the  Levi-Civita connection), and the gravitational field is described by the {\it vierbein }fields  $e^i_\mu(x)$  whose dynamics is governed  by  the torsion tensor (instead of the curvature tensor), that is 
\begin{equation}\label{torsion}
{\cal T}^\lambda_{\mu\nu}=\hat{\Gamma}^\lambda_{\nu\mu}-\hat{\Gamma}^\lambda_{\mu\nu}
=e^\lambda_i(\partial_\mu e^i_\nu - \partial_\nu e^i_\mu)\,.
\end{equation}
Here $e^i_\mu(x)$ are  defined as $g_{\mu\nu}(x)=\eta_{ij} e^i_\mu(x)e^j_\nu(x)$. The possible action is given by
\begin{equation}\label{action}
    S_I = \frac{1}{16\pi G}\int{d^4xe\left[{\cal T}+f({\cal T})\right]},
\end{equation}
where ${\cal T}={S_\rho}^{\mu\nu}{\cal T}^\rho_{\mu\nu}$ is the torsion scalar, $e=det(e^i_\mu)=\sqrt{-g}$, $f({\cal T})$ is a generic function of the torsion,
and
\begin{equation}\label{s}
    {S_\rho}^{\mu\nu} = \frac{1}{2}\left[\frac{1}{4}({{\cal T}^{\mu\nu}}_\rho-{{\cal T}^{\nu\mu}}_\rho-{{\cal T}_\rho}^{\mu\nu})
     + \delta^\mu_\rho
{{\cal T}^{\theta\nu}}_\theta-\delta^\nu_\rho {{\cal T}^{\theta\mu}}_\theta\right]\,.
\end{equation}
TEGR represents an alternative to GR where the coincidence between geodesic structure and causal structure is not required.
This means that Equivalence Principle is not at the foundation of the theory but affine structure  is more relevant. In cosmology, the approach revealed extremely useful to address inflation as well as the late accelerated expansion. In this perspective, the dynamical key role is played by torsion instead of curvature \cite{Ferraro2,Linder:2010py,Pereira.book,Cai:2015emx}). 

For the above  homogeneous,  isotropic and spatially-flat  FLRW Universe,  one finds that $e_{\mu}^A={\rm diag}(1,a,a,a)$ and ${\cal T}=-6H^2$. The cosmological field equations are \cite{Cai:2015emx}
\begin{equation}\label{friedmann}
    12H^2[1+f_{\cal T}]+[{\cal T}+f]=16\pi G\rho,
    \end{equation}
    \begin{equation}
     48H^2f_{{\cal T}{\cal T}}\dot{H}-(1+f_{\cal T})[12H^2+4\dot{H}]-({\cal T}-f)=16\pi G p\,,\nonumber  
\end{equation}
where $f_{\cal T}=df/d{\cal T}$. As an explicit example, we consider the power-law $f(T)$ model \cite{Nesseris:2013jea,fTBBN,Ferraro3}
 \[
 f({\cal T}) = \beta_{\cal T} |{\cal T}|^{n_{\cal T}}\,,
 \]
 where $n_{\cal T}$ has, in principle, another meaning with respect to the above $n$ used in $f(R)$ gravity. By rewriting (\ref{friedmann}) in the form
 \[
H_{TEGR}^2+H_{\cal T}^2=\frac{8\pi}{3M_{Pl}^2}\rho\,,
 \]
where
 \[
H_{\cal T}^2\equiv \frac{f}{6}-\frac{Tf_{\cal T}}{3}=6^{(n_{\cal T}-1)}\beta_{\cal T}(2n_{\cal T}+1)H^{2n_{\cal T}}\,,
 \]
and assuming $H_{\cal T}\gg H_{TEGR}$, one gets the expressions as in (\ref{ATf}) and (\ref{ATGen}), with
 \begin{eqnarray}\label{ATTorsion}
   \eta&=&1\,, \quad \nu=\frac{2}{n_{\cal T}}-2\,, \\
   T_* &\equiv& \left(\frac{24\pi^3 g_*}{45}\right)^{\frac{1}{4}}
   (2n_{\cal T}+1)^{\frac{1}{4(1-n_{\cal T})}}\left(\frac{\beta_{\cal T}}{\mbox{GeV}^{2(1-n_{\cal T})}}\right)^{\frac{1}{4(1-n_{\cal T})}}\left(\frac{M_{Pl}}{\mbox{GeV}}\right)^\frac{1}{2}\mbox{GeV}\,. \nonumber
 \end{eqnarray}
or explicitly (see (\ref{ATTorsion}))
 \begin{equation}\label{A(T)Torsion}
   A(T)=\left(\frac{1}{(2n_{\cal T}+1)}\frac{\text{GeV}^{2(1-n_{\cal T})}}{\beta_{\cal T}}
   \right)^{\frac{1}{2n_{\cal T}}}\left[\left(\frac{45}{24\pi^3 g_*}\right)^{\frac{1}{2}}\left(\frac{T}{\text{GeV}}\right)^2
 \frac{\text{GeV}}{M_{Pl}}\right]^{\frac{1}{n_{\cal T}}-1}\,.
 \end{equation}
In order to solve the gravitino problem, we have to consider the bound $n_{\cal T}\leqslant 2/3$ \cite{fTBBN}, i.e. $\nu\geqslant 1$, as imposed by BBN constraints. We shall discuss the case $n_{\cal T}=2/3$, i.e. $\nu=1$, and $n_{\cal T}<2/3$, i.e. $\nu>1$.

The gravitino problem is studied by means of the Boltzmann (\ref{boltzmann2}). For $\nu=1$ and $\nu>1$, the integration of (\ref{boltzmann2}) from $T_*$ to $T_R$ (assuming $T_R \gg T_*$ and using (\ref{ATTorsion})) gives
 \[
 Y_{3/2}\sim \left\{ \begin{tabular}{ll}  $\frac{T_*}{\eta} \log \frac{T_R}{T_*}$ & $\qquad  \nu=1$  \\
                 & \\
            $\frac{T_*^{2\nu-1}}{\eta(\nu-1)}$ & $\qquad \nu >1$  \end{tabular} \right.
  \]
which is independent of the reheating temperature. Therefore, if the transition temperature is low enough, i.e. $T_* < (10^6 \div 10^7)$GeV, the gravitino problem is avoided even if the reheating temperature is much higher.



\section{Discussion and Conclusions}

Before drawing our conclusions, let us check that the condition $T_R\gg T_*$ occurs also in non-standard cosmology. This result applies to the above models of modified cosmologies. By comparing the expansion rate of the Universe
(\ref{ATf}) with the decay rate of the inflaton $\Gamma_{I}$, one obtains the reheating temperature
 \begin{equation}\label{TRInfla}
   T_R = T_* \left(\frac{M_{Pl}\Gamma_I}{T_*^2}\right)^{\frac{1}{\nu+2}}\,.
 \end{equation}
In the standard cosmology, the relation $H_{GR}(T_R^{GR})=\Gamma_I$ gives
$T_R^{GR}\simeq (M_{Pl}\Gamma_I)^{1/2}$. Combining this relation with the reheating temperature
(\ref{TRInfla}), one infers
 \begin{equation}\label{TRfinal}
   T_R = T_* \left(\frac{T_R^{GR}}{T_*}\right)^{\frac{2}{\nu+2}}\,.
 \end{equation}
Thus $T_R \gg T_*$ provided $T_R^{GR} \gg T_*$. This relation, however, is such that the values $\nu \gg 1$ are excluded. \\

To conclude, we have presented some preliminary results related to the gravitino problem assuming that the background is described by modified  cosmological models coming from Extended Gravity. Using the fact that the expansion rate of the Universe gets modified as $H=A(T)H_{GR}$, where the factor $A(T)$ (given in Eq. (\ref{ATf})) accounts for the non-standard  evolution of the cosmic background, we solved the Boltzmann equation describing the time evolution of the gravitino abundance. It is possible to show that, under specific condition, modified cosmology can provide natural cosmological scenarios able to avoid the late overproduction of gravitinos.
The analysis carried out in this paper relies on models derived from  $f(R)$ and $f({\cal T})$ gravities. However, besides the possibility to find  more general solutions, other curvature and torsion invariants might play a relevant role for the gravitino problem here studied. In particular, cosmological models of the form $f(R, \Box R, \Box^l R, \ldots)$ or   $f({\cal T}, \Box {\cal T}, \Box^l{\cal T }, \ldots)$ deserve to be taken into account \cite{Cap1,Cap2}. Such studies will be considered in  forthcoming papers.


\acknowledgments
The Authors  are supported  by the INFN {\it sezione di Napoli} and {\it gruppo collegato di Salerno}, {\it iniziative specifiche}  MOONLIGHT2 and  QGSKY.


\begin{thebibliography}{0}

\bibitem{wess} J. Wess and B. Zumino, Nucl. Phys. B {70}, 39 (1974).
\bibitem{ferrara} E. Cremmer, S. Ferrara, L. Girardello, and A. van Proyen, Nucl. Phys. B 212 (1983).
\bibitem{ellisN} J. Ellis, C. Kounnas, and D.V. Nanopoulos, Phys. Lett. B 143, 410 (1984). J. Ellis, K. Enqvist, and D.V. Nanopoulos, Phys. Lett. B 147, 99 (1984).
\bibitem{sat} N. Okada and O. Seto, Phys. Rev. D 73, 063505 (2006).
\bibitem{gravitinoPr} M.Y. Khlopov and A. D. Linde, Phys. Lett. B {\bf 138}, 265 (1984).
            J. R. Ellis, J. E. Kim, and D.V. Nanopoulos, Phys. Lett. B {\bf 145}, 181 (1984).
            M. Kawasaki, K. Kohri, and T. Moroi, Phys. Lett. B {\bf 625}, 7 (2005).
\bibitem{moroi} M. Kawasaki, T. Moroi, Progr. Theor. Phys. {\bf 93}, 879 (1995).
\bibitem{olive} R.H. Cyburt, J.R. Ellis, B.D. Fields, and K.A. Olive, Phys. Rev. D {\bf 67}, 103521 (2003).
             M. Kawasaki, K. Khori, and T. Moroi, Phys. Lett. B {\bf 625}, 7 (2005).
\bibitem{linde} K.A. Olive, Phys. Rep. {\bf 190}, 307 (1990).
                D.H. Lyth and A. Riotto, Phys. Rep. {\bf 314}, 1 (1999).
\bibitem{mazumdar} A. Mazumdar, Phys. Rev. D {\bf 64}, 027304 (2001): Nucl. Phys. b {597}, 561 (2001).
                M.C. Bento, N.C. Santos, and R.G. Felipe, Phys. Rev. D {\bf 69}, 123513 (2004).
                N. Okada and O. Seto, Phys. Rev. D {\bf 71}, 023517 (2005); Phys. Rev. D {\bf 73}, 063505 (2006).
\bibitem{accUn} A.G. Reiss {\it et al.}, Astron. J. {\bf 116}, 1009 (1998).
            S. Perlmutter {\it et al.}, Nature (London) {\bf 391}, 51 (1998).
\bibitem{amendolabook} L. Amendola and S. Tsujikawa, {\it Dark Energy: Theory and Observations}, Cambridge University Press, 2010.
\bibitem{odintsovreport} 
S. Capozziello,  Int. J. Mod.Phys. D {\bf11},  483 (2002).
S. Capozziello, M. De Laurentis,  Phys.Rept. {\bf 509}, 167  (2011). 
S. Nojiri and S.D. Odintsov, Phys. Rep. {\bf 505}, 59 (2011).
   A. Silvestri and M. Trodden,  Rep. Prog. Phys. {\bf 72}, 096901 (2009). 
   J.A. Frieman, M.S. Turner, and D. Huterer,  Annu. Rev. Astron. Astrophys. {\bf 46}, 385 (2008).
   R. Durrer and R. Maartens, Gen. Relat. Grav. {\bf 40}, 301 (2008). 
    S. Capozziello and G. Lambiase, Frascati Phys. Ser. {\bf 58}, 17 (2014). 
    M. Sami, Lect.Notes Phys. {\bf 720}, 219 (2007). 
    E.J. Copeland, M. Sami, and Sh. Tsujikawa, Int. J. Mod. Phys. {\bf D} 15, 1753 (2006). 
  T. Clifton, P.G. Ferreira, A. Padilla, and C. Skordis, Phys. Rep. {\bf 513}, 1 (2012).
  T.P. Sotiriou and V. Faraoni, Rev. Mod. Phys. {\bf 82}, 451 (2010).
  A. De Felice and S. Tsujikawa, Living Rev. Rel. {\bf 13}, 3 (2010).
\bibitem{curvquin} S. Nojiri, S.D. Odintsov, 
                            Phys. Rev. D {\bf 77}, 026007 (2008).
            H. Oyaizu, M. Lima, and W. Hu, 
                            Phys. Rev. D {\bf 78}, 123524 (2008).
            L. Pogosian and A. Silvestri, 
                            Phys. Rev. D {\bf 77}, 023503 (2008).
            I. Sawicki and W. Hu, 
                            Phys. Rev. D {\bf 75}, 127502 (2007).
            B. Li and J.D. Barrow, 
                            Phys. Rev. D {\bf 75}, 084010 (2007).
            T. Clifton, 
                            Phys. Rev. D {\bf 78}, 083501 (2008).
%
S. Capozziello and G. Lambiase, Gen. Relat. Grav. {\bf 32}, 295 (2000).
%
S. Capozziello and G. Lambiase, Gen. Relat. Grav. {\bf 31}, 1005 (1999).
%
T. Clifton and J.D. Barrow, 
                            Phys. Rev. D {\bf 72}, 103005 (2005).
 W. Hu and I. Sawicki, 
                            Phys. Rev. D {\bf 76}, 064004 (2007).
 A.A. Starobinsky, JETP Lett. {\bf 86}, 157 (2007).
\bibitem{fornengo}
            S. Capozziello, M. de Laurentis, and G. Lambiase,  Phys. Lett. B {\bf 715}, 1 (2012).
              G. Lambiase, Phys. Rev. D {\bf 90}, 064050 (2014).
      S. Capozziello, G. Lambiase, M. Sakellariadou, An. Stabile, and A. Stabile, Phys.Rev. {\bf D 91}, 044012 (2015).
     G. Lambiase, S. Mohanty, and A.R. Prasanna, Int. J. Mod. Phys. D {\bf 22}, 1330030 (2013).
       G. Lambiase, L. Mastrototaro, Astrophys. J. {\bf 904:19}, 1 (2020).
\bibitem{randal} L. Randal and R. Sundrum, Phys. Rev. Lett. {\bf 83}, 4690 (1991).
\bibitem{kination} F. Profumo and P. Ullio, JCAP {\bf 0311}, 006 (2003).
                F. Rosati, Phys. Lett. B {\bf 570}, 5, (2003).
                C. Pallis, JCAP {\bf 0510}, 015 (2005).
\bibitem{STcosmology} R. Catena, N. Fornengo, A. Masiero, M. Pietroni, and F. Rosati, Phys. Rev. D {\bf 70}, 063519 (2004).
\bibitem{PRDGall} S. Capozziello, V. Galluzzi, G. Lambiase, and L. Pizza, e-Print: arXiv:1507.06835 [astro-ph.CO].
\bibitem{mazPRL} A. Conroy, A. Mazumdar, and A. Teimouri, Phys. Rev. Lett. {\bf 114}, 201101 (2015).


\bibitem{Ferraro2} R. Ferraro and F. Fiorini, Phys. Rev. D. {\bf 75}, 084031 (2007).
 
 \bibitem{Pereira.book} R. Aldrovandi, J.G. Pereira, {\it{Teleparallel Gravity: An Introduction}}, Springer, Dordrecht, 2013.
  
\bibitem{Linder:2010py}
  E.~V.~Linder,
  Phys.\ Rev.\  D {\bf 81}, 127301 (2010).



\bibitem{Cai:2015emx}
  Y.~F.~Cai, S.~Capozziello, M.~De Laurentis and E.~N.~Saridakis,
  Rept.\ Prog.\ Phys.\  {\bf 79}, no. 10, 106901 (2016).

\bibitem{Nesseris:2013jea}  S.~Nesseris, S.~Basilakos, E.~N.~Saridakis and L.~Perivolaropoulos,  Phys.\ Rev.\ D {\bf 88}, 103010 (2013).
\bibitem{fTBBN} S. Capozziello, G. Lambiase, E.N. Saridakis, Eur. Phys. J. C {\bf 77}, 576 (2017).
        M. Benetti, S. Capozziello, G. Lambiase, Mon. Not. Roy. Astron. Soc.   {\bf  500}, 1795 (2021).
\bibitem{Ferraro3} R. Ferraro and F. Fiorini, Phys. Rev. D {\bf 78}, 124019 (2008).
\bibitem{FRWshear} G.F.R. Ellis, M.A.H. MacCallum, Commun. Math. Phys. 12 (1969) 108.
    M.A.H. MacCallum, G.F.R. Ellis, Commun. Math. Phys. 19 (1970) 31.
    M.P. Ryan, L.C. Shepley, {\it Homogenous Relativistic Cosmologies}, Princeton University Press, Princeton, NJ, 1975.
\bibitem{ale} H. Iminniyaz, B. Salai, and G.-L. Lv, Commun. Theor. Phys. Vol. {\bf 70}, 602 (2018).
            A. Poulin, Phys. Rev. D {\bf 100}, 043022 (2019).
            
\bibitem{Cap1}
  S.~Capozziello, M.~Capriolo and L.~Caso,
  Class.\ Quant.\ Grav.\  {\bf 37} ,   235013 (2020).
  
\bibitem{Cap2}
  S.~Capozziello, M.~Capriolo and S.~Nojiri,
  Phys.\ Lett.\ B {\bf 810},  135821 (2020).
  
  \end{thebibliography}
\end{document}